\documentclass[10pt,twocolumn,letterpaper]{article}

\usepackage{cvpr}              

\usepackage{graphicx}
\usepackage{amsmath}
\usepackage{amssymb}
\usepackage{booktabs}
\usepackage{soul} 

\usepackage{color}

\usepackage{caption} 
\usepackage{graphicx}
\usepackage{colortbl}
\usepackage{multirow}
\usepackage{xspace}
\usepackage{float}
\usepackage[T1]{fontenc} 
\usepackage[utf8]{inputenc} 

\newcommand{\rnet}{\textit{ResNet18}\xspace}
\newcommand{\dtt}{\textit{DeiT-Tiny}\xspace}

\newcommand{\pnet}{\textit{PathNet}\xspace}
\newcommand{\dino}{\textit{DINO}\xspace}

\usepackage[pagebackref,breaklinks,colorlinks]{hyperref}

\usepackage[capitalize]{cleveref}
\crefname{section}{Sec.}{Secs.}
\Crefname{section}{Section}{Sections}
\Crefname{table}{Table}{Tables}
\crefname{table}{Tab.}{Tabs.}

\def\cvprPaperID{41} 
\def\confName{CVPR}
\def\confYear{2022}

\title{\vspace{-1cm}A comparative study between vision transformers and CNNs in digital pathology}
\author{Luca Deininger$^1$\\[-2pt]
{\tt\footnotesize deininger.luca@gmail.com}
\and
Bernhard Stimpel$^1$\\[-2pt]
{\tt\footnotesize bernhard.stimpel@roche.com}
\and
Anil Yuce$^1$\\[-2pt]
{\tt\footnotesize anil.yuce@roche.com}
\and
Samaneh Abbasi-Sureshjani$^1$\\[-2pt]
{\tt\footnotesize samaneh.abbasi@roche.com}
\and
Simon Sch\"onenberger$^1$\\[-2pt]
{\tt\footnotesize simon.schoenenberger@gmail.com}
\and
Paolo Ocampo$^2$\\[-2pt]
{\tt\footnotesize ocampo.paolo-santiago@gene.com}
\and
Konstanty Korski$^1$\\[-2pt]
{\tt\footnotesize konstanty.korski@roche.com}
\and
Fabien Gaire$^1$\\[-2pt]
{\tt\footnotesize fabien.gaire@roche.com}}
\date{%
    $^1$ F. Hoffmann-La Roche AG, Grenzacherstrasse 124, 4070 Basel, Switzerland\\%
    $^2$ Genentech, Inc., 1 DNA Way, South San Francisco, CA 94080, USA\\[2ex]%
}

\usepackage{graphicx}

\begin{document}
\maketitle

\begin{abstract}
Recently, vision transformers were shown to be capable of outperforming convolutional neural networks when pretrained on sufficient amounts of data. In comparison to convolutional neural networks, vision transformers have a weaker inductive bias and therefore allow a more flexible feature detection. Due to their promising feature detection, this work explores vision transformers for tumor detection in digital pathology whole slide images in four tissue types, and for tissue type identification. We compared the patch-wise classification performance of the vision transformer \dtt to the state-of-the-art convolutional neural network \rnet. Due to the sparse availability of annotated whole slide images, we further compared both models pretrained on large amounts of unlabeled whole-slide images using state-of-the-art self-supervised approaches. The results show that the vision transformer performed slightly better than the \rnet for three of four tissue types for tumor detection while the \rnet performed slightly better for the remaining tasks. The aggregated predictions of both models on slide level were correlated, indicating that the models captured similar imaging features. All together, the vision transformer models performed on par with the \rnet while requiring more effort to train. In order to surpass the performance of convolutional neural networks, vision transformers might require more challenging tasks to benefit from their weak inductive bias.
\end{abstract}

\section{Introduction}
\label{sec:intro}

Convolutional neural networks~(CNNs) are a powerful tool for tumor detection in digital pathology whole-slide images~(WSIs) reaching accuracies of over 90\% for patch-wise tumor detection depending on the type of cancer~\cite{talo2019, KHAN20191, Tabibu2019, VO2019123}.
CNNs have been the common architecture for digital pathology tasks, but the more recently introduced vision transformers~(ViTs)~\cite{2010.11929} have not been fully investigated in this domain yet. In comparison to CNNs, ViTs are free of convolution-induced biases which allows the model to learn global features and complex relations in the data. Dosovitskiy \etal~\cite{2010.11929} showed that their entirely convolution-free ViT can outperform CNNs such as ResNet~\cite{he2015deep}. Also considering the strength of transformers for natural language processing~\cite{1810.04805, 1906.08237, Radford2019LanguageMA, 1907.11692}, this raises the question whether ViTs are the next big model in image recognition and whether those will replace CNNs in the future. 

Despite the triumph of ViTs, their weak inductive bias is also a limiting factor. Meaning, those models require large amounts of training data to surpass the performance of CNNs~\cite{2010.11929, 2012.12877}. Touvron \etal~\cite{2012.12877} aimed for training ViTs on datasets of the same size as comparable CNNs. They introduced the data-efficient vision transformer tiny~(\dtt), which we focus on in this work.
Beyond architectural model adjustments, transfer learning  can overcome the difficulty of training vision transformers as it improves model convergence for downstream tasks~\cite{ravishankar2017understanding}. In digital pathology, models pretrained on ImageNet~\cite{deng2009imagenet} are common. However, more recent work showed the potential of models pretrained on digital pathology datasets~\cite{pmlr-v156-abbasi-sureshjani21a}. Especially self-supervised pretraining is interesting in this domain due to the sparse availability of annotated WSIs. This allows the model to learn features on large amounts of data, which might boost the performance for downstream tasks on unseen datasets.
A recent promising self-supervised approach for ViT training is
DINO~\cite{2104.14294}. The authors of DINO present their method as self-distillation with no labels and showed that the extracted self-supervised DINO features are powerful for other downstream image classification tasks. Even though their method is not restricted to ViTs, the authors demonstrate that the DINO architecture works especially well for those models.

Due to their emerging potential, this work explores fully- and self-supervised~(DINO) ViTs for patch-wise tumor detection in sentinel lymph node~(SLN), diffuse large B-cell lymphoma~(DLBCL), breast, and lung adenocarcinoma~(LUAD) WSIs. Furthermore, we benchmarked the models on tissue type identification in colorectal cancer WSIs. To evaluate the potential of ViTs, we compared their performance to the state-of-the-art CNN \rnet and analyzed the qualitative differences in their classifications.

\section{Materials and Methods}
This section presents the configuration of the fully- and self-supervised ViT and the setup of the baseline \rnet. Moreover, it explains the datasets used for the two downstream tasks tumor detection and tissue type identification.

\subsection{Fully-supervised models} 

\subsubsection{Model configuration}
In this work, we used the \rnet pretrained on ImageNet as the baseline classifier, as it provides a good trade-off between classification accuracy and training time for the tasks considered in this work. For the ViT, we used the data-efficient image transformer tiny~(\dtt) with patch size 16, pretrained on ImageNet~\cite{2012.12877}.

For both models, we changed the number of neurons in the last fully-connected layer to the number of target classes.

\subsubsection{Model training}
\label{sec:fs_training}
The fully-supervised models used the \textit{Cross-entropy Loss} weighted according to inverse class frequencies and trained for up to 300 epochs using batch size 128. The \rnet based models used \textit{Adam}~\cite{1412.6980} with $\beta_1=0.9$, $\beta_2=0.98$, learning rate $1e^{-6}$, and L2 regularization~($1e^{-3}$). The \dtt used \textit{sharpness-aware minimization}~(SAM)~\cite{2010.01412} using the implementation from Samuel~\cite{sam_implementation} with stochastic gradient descent~(SGD), learning rate $1e^{-3}$ and momentum $0.9$. The idea of SAM is to simultaneously minimize the loss value and the loss sharpness. It seeks model parameters that lie in neighborhoods having a low loss. For this purpose, SAM performs two forward-backward passes for gradient descent which results in a longer training time in comparison to Adam. Initially, we used Adam for ViT optimization but we observed that the model did not achieve the same classification performance as the \rnet. SAM greatly improved ViT generalization in comparison to Adam at the cost of an increased runtime.

To speed up the training time per epoch, the models used balanced sampling of the training patches and stratified subsampling of the validation patches, resulting in roughly 100,000 training and 50,000 validation patches per epoch. 

In order to prevent overfitting, the models used an \textit{albumentations}~(version 0.5.2)~\cite{2018arXiv180906839B} image augmentation pipeline comprising \textit{ColorJitter}, vertical and horizontal flip, and 90 degree rotation. Furthermore, the models used early stopping with 30 epochs patience to prevent overfitting.

\subsection{Self-supervised models}
For self-supervised ViT pretraining, we utilized DINO~\cite{2104.14294} with a \dtt backbone and trained it on what we call the TCGA 100 dataset. We refer to this model as \dino in the following. The TCGA 100 dataset comprises patches from 8,747 \textit{The Cancer Genome Atlas}~(TCGA) WSIs encompassing the datasets BRCA, CHOL, HNSC, KIRC, KIRP, LIHC and PRAD. These slides come from around 20 different tissue types. The models randomly sampled 100,000 training and 50,000 validation patches at each epoch to reduce the training time per epoch, and trained for 100 epochs. Furthermore, the models used $256 \times 256$ images instead of $224 \times 224$ images as in the original DINO implementation.

For transfer to tumor detection and tissue type identification, we used the \dino teacher as a \dtt backbone and trained it in a fully-supervised manner as explained in~\cref{sec:fs_training}. In addition to this setup, we trained the neural network classifier only and fixed the weights in the feature extraction layers to speed up downstream training. We refer to this setting as FW~('fixed weights') in the following sections. For tissue type identification, we further used image up-scaling to $256 \times 256$ to fit the pretrained \dino backbone architecture.

As a self-supervised CNN, we used the \rnet pretrained in a self-supervised manner on a digital pathology dataset, termed as \pnet~\cite{pmlr-v156-abbasi-sureshjani21a}. \pnet was trained with \textit{Bootstrap Your Own Latent}~(BYOL)~\cite{2006.07733} on eight tissue types: DLBCL, lymph node, follicular lymphoma, tonsil, lung, colon, breast and thyroid. The reader is referred to Abbasi-Sureshjani~\etal~\cite{pmlr-v156-abbasi-sureshjani21a} for details on model pretraining. For transfer to downstream tasks, we trained it as described in~\cref{sec:fs_training}.

\subsection{Performance evaluation}
In order to assess the model performance for tumor detection, we calculated the area under the precision-recall curve~(PR~AUC). For tissue type identification, we macro averaged the PR AUC for each class~(one vs. rest) to retrieve one value. Since accuracy is the most frequently reported metric in the literature, we also report it for the comparison of our models to state-of-the-art approaches.

\subsection{Model comparison}
In order to assess the similarity of \rnet and \dtt predictions, we calculated the Pearson correlation~\cite{freedman2007statistics} of the mean patch-wise accuracy of the test slides. Moreover, we computed a two-sided p-value which indicates the probability of uncorrelated model predictions that have a Pearson correlation at least as extreme as the one computed~\cite{2020SciPy-NMeth}. For attention heatmap creation, we used \textit{Gradient-weighted Class Activation Mapping} (\textit{Grad-CAM})~\cite{1610.02391}. Our implementation used the last feature extraction layer as the model target layer and backpropagated the label of the model's own prediction.

\subsection{Datasets}

\subsubsection{Tissue type identification}
For tissue type identification we used the CRC9 dataset~\cite{kather_jakob_nikolas_2018_1214456} comprising 100,000 non-overlapping $224 \times 224$ color-normalized~\cite{Macenko2009} patches at magnification $20\times$~(0.5 microns per pixel~[MPP]). The patches originate from hematoxylin~\&~eosin~(H\&E) stained histological images of human colorectal cancer~(CRC) and normal tissue, and are categorized in nine classes. Our models used 70\% of the data for training, 15\% for validation and 15\% for testing.

\subsubsection{Tumor detection}
For all tumor detection tasks, we used our internal tissue segmentation software to determine the WSI tissue region. Patches belonging to the tumor class were obtained through exhaustive~(all tumor lesions in a slide were annotated) and non-exhaustive pathologist tumor annotations. The patches belonging to the non-tumor~(normal) class were obtained either from WSIs of tissues known to contain no tumor lesions, or from regions outside the pathologist exhaustive tumor lesion annotations. If not otherwise mentioned, we extracted $256 \times 256$ patches at magnification $20\times$~(0.5 MPP) with an overlap of at least 90\% with the desired region~(tumor or normal) and sampled 75\% of the slides for training and 25\% for validation while using an independent test set.

We benchmarked our models on four tissue types. We used the Camelyon16 dataset~\cite{10.1001/jama.2017.14585} for SLN metastatic tumor detection which comprises 270 training and 130 test WSIs from two independent medical centers. For DLBCL tumor detection, we trained and tested on 4,957 and 103 internal WSIs, respectively. Those WSIs comprised non-tumoral tonsil and lymph node tissue, and lymph node tissue with DLBCL tumor. For breast tumor detection, we trained and tested on 335 and 537 internal breast WSIs, respectively. For LUAD tumor detection, we trained and tested on 431 and 138 internal lung WSIs, respectively. For details, see Supplementary.
\vspace{-0.1cm}
\section{Results}\vspace{-0.1cm}
In this section, we present the results for patch-wise tissue type identification, and tumor detection for several tissue types. We compared the performance of the CNNs \rnet and \pnet to the ViTs \dtt and \dino. Furthermore, we compared their classifications on slide level.

\subsection{Performance comparison across all datasets}
The comparison of the test PR AUC values across all datasets, shown in \cref{tab:all_test_performances}, demonstrates that the performance of the \rnet, \pnet, the \dtt and \dino were very similar. For DLBCL, the \rnet performed slightly better while the ViT-based models performed slightly better for SLN, LUAD and breast.

\begin{table}[H]
\footnotesize
\centering
\caption{Model test PR AUC and accuracy (ACC). For datasets with multiple test sets~(LUAD and breast), we show the unweighted mean per dataset.}%
\label{tab:all_test_performances}%
\addtolength{\tabcolsep}{-2.5pt}
\begin{tabular}{lcl|rrrrr}
\hline
Model& FW& Metric & \multicolumn{1}{c}{CRC9} & \multicolumn{1}{c}{SLN} & \multicolumn{1}{c}{DLBCL} & \multicolumn{1}{c}{LUAD} & \multicolumn{1}{c}{Breast} \\ \hline
\multirow{2}{*}{\rnet} & \multirow{2}{*}{×} & PR AUC & \textbf{0.999}& 0.885& \textbf{0.976} & 0.913& 0.809  \\
  && ACC    & \textbf{0.995}& 0.981& \textbf{0.88}  & 0.858& 0.915  \\ \hline
\multirow{2}{*}{\dtt}  & \multirow{2}{*}{×} & PR AUC & 0.998& \textbf{0.917}& 0.97  & \textbf{0.94} & 0.817  \\
  && ACC    & 0.982& \textbf{0.988}& 0.874 & 0.88 & 0.913  \\ \hline
\multirow{2}{*}{\pnet} & \multirow{2}{*}{×} & PR AUC & \textbf{0.999}& 0.908& 0.97  & 0.92 & 0.818  \\
  && ACC    & \textbf{0.995}& 0.979& 0.866 & \textbf{0.885}& 0.92   \\ \hline
\multirow{2}{*}{\dino} & \multirow{2}{*}{×} & PR AUC & \textbf{0.999}& 0.912& 0.958 & 0.933& \textbf{0.828}  \\
  && ACC    & 0.991& 0.984& 0.874 & 0.871& \textbf{0.924}  \\ \hline
\end{tabular}
\end{table}

During early experiments with SLN tumor detection, we assessed whether the \dtt benefits from more WSI context. Thus, we trained and tested the models on $10\times$ magnification. The performance was very similar to training on $20\times$. On $10\times$, the \rnet and the \dtt reached a test PR AUC of $0.887$ and $0.919$, respectively.

To place our results in context to existing work, we compared our results to state-of-the-art methods on the same digital pathology tasks. All together, our results are on par with the state-of-the-art performances on the same digital pathology tasks~\cite{Kather2019, Steinbuss2021, WANG2019103, Hatuwal2020, Zhu2019,Mi2021}. However, the results are not directly comparable due to differences in the datasets and test splits used.

\subsection{\textbf{\pnet} vs. \textbf{\dino}}
We observed that when fixing the weights in the feature extraction layers and using the extracted features directly, \pnet performed better than \dino for DLBCL only~(\cref{tab:pathnet_vs_dino_tcga_across_datasets_fw}). For CRC9, SLN, LUAD and breast, \dino performed better which we hypothesize is explained by the different dataset that it was trained on. TCGA 100 contains more diverse tissue types thus the network has learned a better phenotype representation. This suggests to consider \dino for embedding extraction for downstream tasks, since it is more versatile than \pnet.

\begin{table}[H]
\centering
\footnotesize
 \caption{Comparison of \pnet and \dino test PR AUC and accuracy for fixing weights in model feature extraction layers.}
 \label{tab:pathnet_vs_dino_tcga_across_datasets_fw}%
\addtolength{\tabcolsep}{-2pt}
\begin{tabular}{lcl|rrrrr}
\hline
Model& FW& Metric & \multicolumn{1}{c}{CRC9} & \multicolumn{1}{c}{SLN} & \multicolumn{1}{c}{DLBCL} & \multicolumn{1}{c}{LUAD} & \multicolumn{1}{c}{Breast} \\ \hline
\multirow{2}{*}{\pnet} & \multirow{2}{*}{\checkmark} & PR AUC & 0.969& 0.833&\textbf{0.931}&0.887& 0.711  \\
  && ACC    & 0.928& 0.943&\textbf{0.829}&  0.859& 0.883 \\ \hline

\multirow{2}{*}{\dino} & \multirow{2}{*}{\checkmark} & PR AUC & \textbf{0.997}& \textbf{0.894}& 0.887 & \textbf{0.914}& \textbf{0.751}  \\
  && ACC    & \textbf{0.983}&\textbf{0.959}&0.817&\textbf{0.878} & \textbf{0.897} \\ \hline
\end{tabular}
\end{table}

\subsection{Model comparison}

\begin{figure}[htbp]
        \centering
        \begin{subfigure}[b]{0.235\textwidth}
            \centering
            \includegraphics[width=\textwidth]{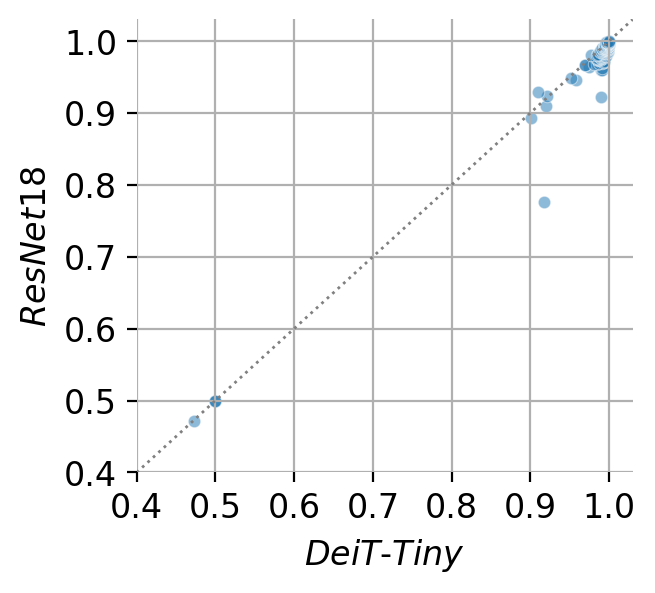}
            \caption{SLN: 0.986~($p=2e^{-100}$)}    
        \end{subfigure}
        \hfill
        \begin{subfigure}[b]{0.235\textwidth}  
            \centering 
            \includegraphics[width=\textwidth]{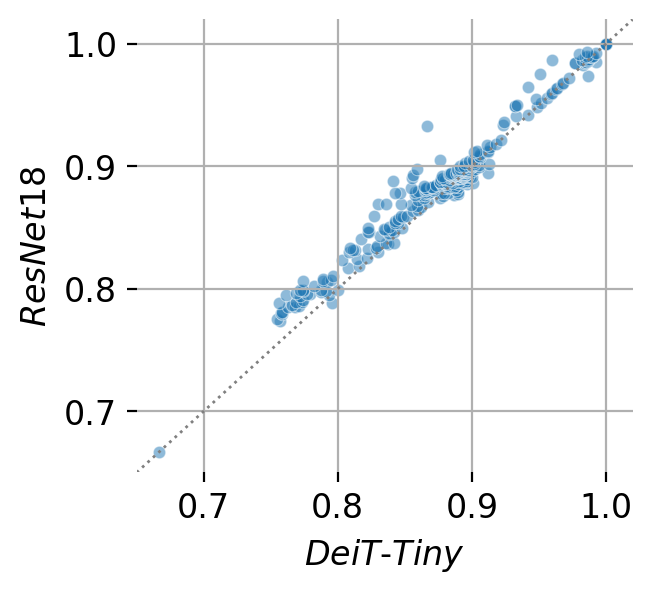}
            \caption{DLBCL: 0.969~($p=2e^{-184}$)}    

        \end{subfigure}
        \begin{subfigure}[b]{0.235\textwidth}   
            \centering 
            \includegraphics[width=\textwidth]{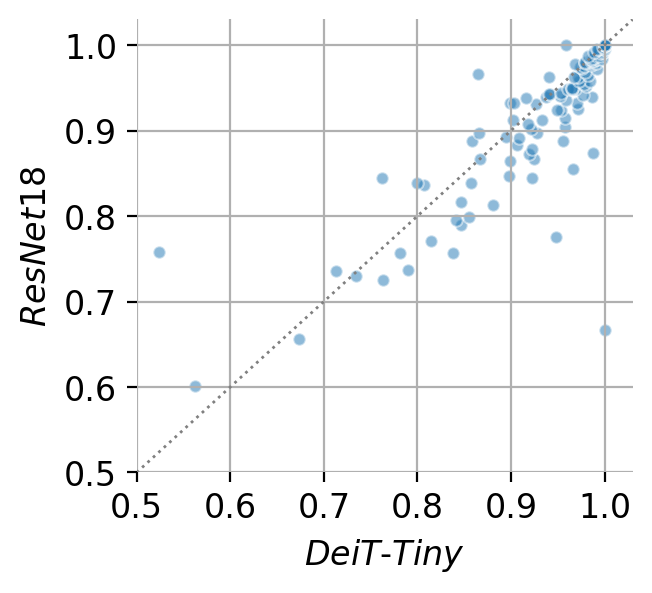}
            \caption{Breast: 0.82~($p=7e^{-28}$)}    
        \end{subfigure}
        \hfill
        \begin{subfigure}[b]{0.235\textwidth}   
            \centering 
            \includegraphics[width=\textwidth]{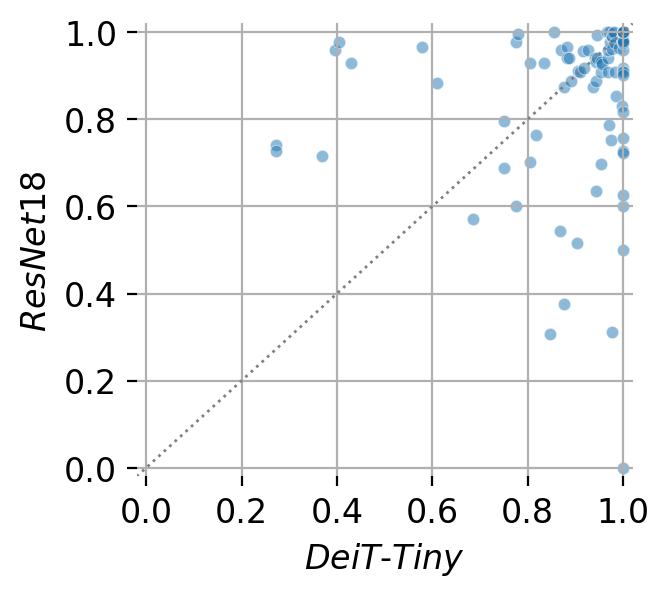}
            \caption{LUAD: 0.1~($p=0.29$)}    
        \end{subfigure}
        \caption{Mean accuracy per test slide incl. Pearson correlation.}
        \label{fig:mean_accuracy_per_test_slide_all_datasets}
    \end{figure}

In order to compare the \rnet and \dtt predictions, we calculated the mean patch-wise accuracy of the test slides for every dataset. The comparison shows a correlation of the mean test slide accuracies for SLN, DLBCL and breast with exception of a few outliers~(\cref{fig:mean_accuracy_per_test_slide_all_datasets}). For LUAD, we hypothesize that mainly smaller slides cause the larger accuracy fluctuations.

\begin{figure}[H]
\centering
\includegraphics[width=0.60\linewidth]{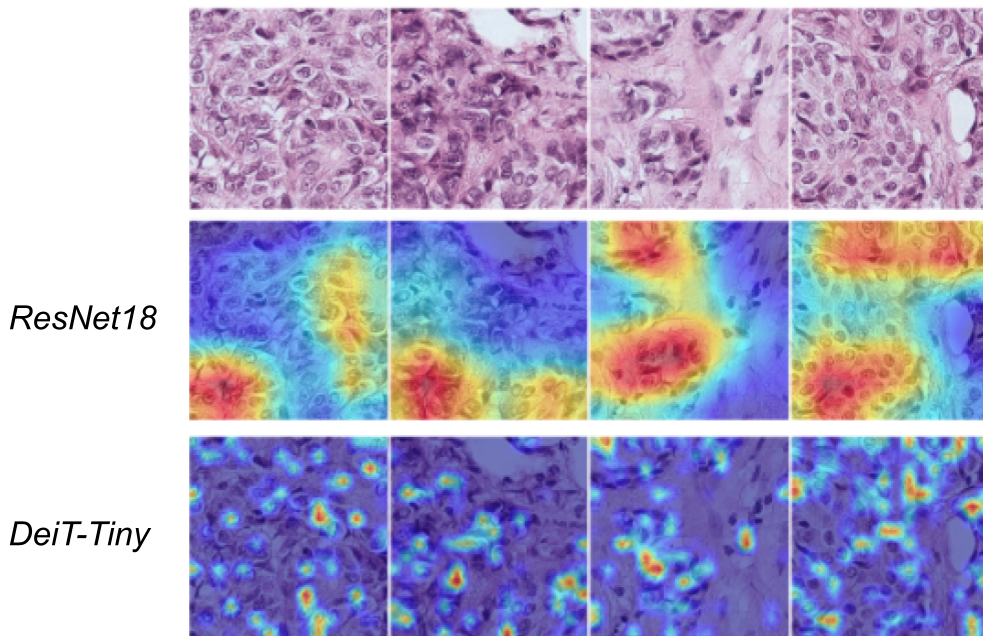}

\caption{Comparison of \rnet and \dtt \textit{Grad-CAM} heatmaps for randomly selected SLN tumor patches. Both models classified all shown patches correctly with a probability of $> 0.9$.}
\label{fig:cam16_attention_random_tiles}
\end{figure}

A comparison of \rnet and \dtt \textit{Grad-CAM} heatmaps shows that the ViT focused on more localized patch regions in comparison to the CNN~(\cref{fig:cam16_attention_random_tiles}). While the benefit is not evident for tumor detection, this might be important for other tasks in digital pathology.

\subsection{Runtime comparison}
As \cref{tab:runtime_comparison_models} shows, the \dtt throughput is twice as slow as the \rnet throughput due to SAM which needs to perform two forward-backward passes for training.

\begin{table}[htbp]
\footnotesize
\centering
\caption{Throughput comparison of \rnet and \dtt for full training pipeline.}
\label{tab:runtime_comparison_models}
\begin{tabular}{lrr}
\hline
\multicolumn{1}{l}{Model} & \multicolumn{1}{c}{\begin{tabular}[c]{@{}c@{}}\#params\end{tabular}} & \multicolumn{1}{r}{\begin{tabular}[c]{@{}c@{}}Throughput~(img/s)\end{tabular}} \\ \hline
\rnet & 11M & 250 \\
\dtt~(Adam) & 5M & 238 \\
\dtt~(SAM) & 5M & 117 \\ \hline
\end{tabular}
\end{table}

\vspace{-0.5cm}
\section{Conclusion}

ViTs show emerging potential in the field of image classification due to their weak inductive bias in comparison to CNNs which allows them to understand complex relationships in the data. To assess whether this also holds for tasks in digital pathology, this work explored fully- and self-supervised ViTs for tissue type identification and tumor detection in WSIs.

The benchmark on several datasets showed that the ViT-based models performed similar to the baseline \rnet across several datasets and across different magnifications. Latter observation suggests that neither the \rnet nor the \dtt showed a benefit from more WSI context. The comparison of the classifications on slide level showed that the predictions of the \rnet and ViT-based models are correlated. This indicates that both the \rnet and the ViT captured similar image features. Interestingly, we observed that \dino is more versatile than \pnet for tumor detection across several tissue types and thus offers an interesting alternative to the \pnet backbone. However, the higher performance of \dino is most likely caused by a more diverse dataset trained on.

Considering the similar performances across several datasets and magnifications, the correlation of slide-wise accuracies and the costly ViT training, the \dtt performed on par with the \rnet while requiring more training effort. We hypothesize that digital pathology tissue type identification and tumor detection are tasks that can be easily learned with traditional CNN approaches. Consequently, ViTs often could not benefit from its larger flexibility in feature detection in comparison to the CNNs. Therefore, we propose the application of ViTs to more challenging digital pathology tasks that for example require more contextual knowledge.
\vspace{-0.2cm}
\subsubsection*{Acknowledgements}\vspace{-0.17cm}
The authors would like to thank Roche Diagnostic Solutions, Roche Life Cycle and the Roche Study team for providing the data, as well as the Roche Personalized Healthcare Digital Pathology Program for funding this work. Some of the results presented here are based upon data generated by the TCGA Research Network: \url{https://www.cancer.gov/tcga}. The authors declare the following competing interests:  A.Y., B.S., S.A., F.G. and K.K. are employees of Roche, L.D. and S.S. were employed by Roche at the time of this work and P.O. is employee of Genentech.

{\small
\bibliographystyle{ieee_fullname}
\bibliography{egbib}
}

\end{document}


\maketitle
\section{Dataset details}
\label{sec:appendix_overview_datasets}
\noindent The following tables show an overview of the datasets used for colorectal cancer (CRC) tissue type identification (\cref{tab:crc9_dataset}), sentinel lymph node~(SLN,~\cref{tab:cam16_number_of_extracted_tiles}), diffuse large B-cell lymphoma~(DLBCL,~\cref{tab:dlbcl_datasets}), breast (\cref{tab:breast_datasets}) and lung adenocarcinoma~(LUAD,~\cref{tab:luad_datasets}).

\begin{table}[H]
\centering
\caption{CRC9 dataset: number of patches for training and testing.\\~~~~Number of patches for training include patches for validation.}
\label{tab:crc9_dataset}
\begin{tabular}{lrr}
\hline
Class     & \multicolumn{1}{l}{Training} & \multicolumn{1}{l}{Test} \\ \hline
TUM  & 12,169                        & 2,148                     \\ 
MUS  & 11,506                        & 2,030                     \\ 
LYM  & 9,823                         & 1,734                     \\
DEB  & 9,785                         & 1,727                     \\ 
BACK & 8,981                         & 1,585                     \\ 
STR  & 8,879                         & 1,567                     \\
ADI  & 8,846                         & 1,561                     \\ 
MUC  & 7,562                         & 1,334                     \\ 
NORM & 7,449                         & 1,314                     \\  \hline
\end{tabular}
\end{table}

\begin{table}[H]
\centering
\caption{SLN: number of extracted patches.}
\label{tab:cam16_number_of_extracted_tiles}
\begin{tabular}{llrr}
\hline
   \multicolumn{1}{c}{Magnification} &  Class  & \multicolumn{1}{c}{Training} & \multicolumn{1}{c}{Test} \\ \hline
\multirow{2}{*}{$20\times$}&   Normal       & 2,520,477                    &  1,312,918                    \\ 
 & Tumor           &60,719                   & 52,506                                   \\ \hline
\multirow{2}{*}{$10\times$}& Normal & 638,555                       & 333,719                   \\ 
 & Tumor  & 15,035                        & 13,081                    \\ \hline
\end{tabular}
\end{table}

\begin{table}[H]
\caption{DLBCL datasets. DLBCL train and test are internal datasets.}
\label{tab:dlbcl_datasets}
\centering
\begin{tabular}{lrrrr}
\hline
Dataset  & \multicolumn{1}{l}{\# slides} & \multicolumn{1}{c}{\# normal patches} & \multicolumn{1}{c}{\# tumor patches}  \\ \hline
Camelyon16 Train & 248 & 2,520,477 & - \\ 
Camelyon16 Test & 128 & 1,312,918 & - \\ 
DLBCL train & 4,581 & 30,333,616 & 13,923,512 \\
DLBCL test & 103 & 145,693 & 479,818  \\ \hline
Total & 5,060 & 34,312,704 & 13,923,512 \\ \hline
\end{tabular}
\end{table}

\thispagestyle{empty}
\begin{table}[H]
\centering
\caption{Breast cancer datasets. Non TCGA datasets are internal datasets.}
\label{tab:breast_datasets}
\begin{tabular}{lrrr}
\hline
\multicolumn{1}{l}{Dataset} & \multicolumn{1}{c}{\# slides} & \multicolumn{1}{c}{\# normal patches} & \multicolumn{1}{c}{\# tumor patches} \\ \hline
TCGA train & 167 & 84,471 & 616,963 \\
Breast train & 168 & 30,423 & 154,243 \\
TCGA test 2 & 64 & 37,270 & 222,789 \\
Breast test 2& 45 & 13,681 & 37,037 \\ 
Breast test 1 & 333 & 25,138 & 675,864 \\
TCGA test 1 & 95& 67,139& 369,991\\
\hline
Total & 872	& 258,122&	2,076,887 \\ \hline
\end{tabular}
\end{table}

\begin{table}[H]
\caption{LUAD datasets.}
\label{tab:luad_datasets}
\centering
\begin{tabular}{lrrr}
\hline
Dataset & \# slides & \# normal patches & \# tumor patches \\ \hline
Lung train & 382 & 104,750 & 255,455 \\
TCGA\_LUAD train& 49 & 45,905 & 176,650 \\ 
Lung test 1 & 94& 45,748 & 91,107 \\
Lung test 2& 44 & 31,160 & 40,063 \\ \hline
Total & 569& 227,563 & 563,275 \\ \hline
\end{tabular}
\end{table}

\includegraphics[width=0.60\linewidth]{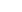}